\documentclass[twocolumn,showpacs,preprintnumbers,amsmath,amssymb,showkeys]{revtex4}

\usepackage{graphicx}
\usepackage{dcolumn}
\usepackage{bm}
\usepackage{bezier}

\newcommand{\PRL}{\textit{Phys. Rev. Lett.}}

\begin{document}

\title{Quasiparticle agglomerates in Read-Rezayi and anti Read-Rezayi 
states}
\author{A. Braggio$^1$, D. Ferraro$^{1,2,3}$ and N. Magnoli$^{2,3}$}
\affiliation{
$^1$ CNR-SPIN, Dipartimento di Fisica, Via Dodecaneso 33, 16146, Genova, Italy\\
$^2$ Dipartimento di Fisica, Universit\`a di Genova,Via Dodecaneso 33, 16146, Genova, Italy\\
$^3$ INFN, Dipartimento di Fisica, Via Dodecaneso 33, 16146, Genova, Italy }%
\date{\today}

\begin{abstract}
We calculate the dominant excitations for the $k$-level ($k\in\mathbb{N}$) Read-Rezayi (RR) states and their particle-hole 
conjugates, the anti Read-Rezayi ($\overline{\textrm{RR}}$),  
proposed for quantum Hall states. These states are supposed to be build over the second Landau level with total 
filling factor $\nu=2+\nu^*$ with $\nu^*=k/(k+2)$ for RR and 
$\nu^*=2/(k+2)$ for $\overline{\textrm{RR}}$. In the $k$-level RR states, based on 
$\mathbb{Z}_k$ parafermions, the dominant excitations are the fundamental quasiparticles with fractional charge $e^*_k= e/(k + 2)$, with $e$ the electron charge, if $k=2,3$. For $k=4$ the single-qp and the $2$-agglomerate, with charge $2e^*_k$, 
have the same scaling and both dominate, while for $k>4$ the $2$-agglomerates are dominant. 
Anyway the dominance of the $2$-agglomerates can be affected by the presence of environmental renormalizations.
For all the $k$-level $\overline{\textrm{RR}}$ states 
the single-qp and the $2$-agglomerate have the same scaling and both dominate. In this case only the presence of 
environmental renormalizations can make one dominant over the other. We determine the conditions where 
the environmental renormalizations 
of the charged and neutral modes make dominant the Abelian $2$-agglomerates over the non-Abelian single-quasiparticles in the two models
and for any value of $k$. We conclude 
observing that, according these predictions, the dominance of $2$-agglomerates, at very low energies for the $\nu=5/2$, can be an interesting indication supporting 
the validity of the anti-Pfaffian model in comparison to the Pfaffian. 
\end{abstract}
\maketitle

\section{Introduction}
\label{sec:Introduction}
Fractional quantum Hall systems are an unique platform in condensed matter physics to study the peculiar properties of low dimensional electron 
systems. The two dimensional nature of the electron gas opens the 
possibility to explore a richer class of electron liquids with exotic excitations and intriguing statistical properties. Quasiparticles (qps) with 
fractional charge, and consequently fractional statistics (neither bosonic neither fermionic), were found.\cite{Laughlin83,dePicciotto97,Saminadayar97} 
The fractional statistics can be not only Abelian but also non-Abelian\cite{Wen91,Read99} such as for the Pfaffian 
  or anti-Pfaffian models developed for $\nu=5/2$.\cite{Fendley07,Levin07,Lee07}\\
  Low-energy effective theories for the edges states have been demonstrated successfull to derive 
 transport and noise properties in the simplest testing device: the quantum point contacts (QPCs).\cite{Chang03} In the Laughlin 
  sequence $\nu=1/(2n+1)$ with $n\in \mathbb{N}$, the gapless modes at the edges can be described in terms of Chiral Luttinger Liquid 
  ($\chi$LL) with minimal excitations of charge $e^*=e/(2n+1)$ where $e$ is the electron charge.\cite{Wen90} The effective edge description for the 
  Jain sequence $\nu=p/(2np+1)$, with $p\in \mathbb{Z}$, were obtained within the hierarchical 
  models where the minimal charge is 
  $e^*=e/(2np+1)$.\cite{Jain89,Wen92} In such cases for $|p|>1$ the hierarchy predicts the presence of $|p|$ 
channel (one charged and $|p|-1$ neutral) with an hidden SU($|p|$) symmetry. 
The experimental observations of shot noise in the QPC at extremely 
  weak backscattering confirmed the value of the fundamental charges supporting the validity of 
  previous models.\cite{Kane94,dePicciotto97,Saminadayar97,Reznikov99} 
  
 More recently, at the lowest possible temperatures, unexpected increasing of the carrier charges were 
 reported. For the Jain's series with $|p|>1$ (such as for example  $\nu=2/5,2/3$) the carrier charge grew up to $\nu e$.\cite{Chung03,Bid09}  
 For $\nu=2/5$ ($p=2$), at low enough energy,
the dominant excitation is 
the $|p|$-agglomerate of  qps with charge $|p|e^*=\nu e$ instead of the single-qp
charge $e^*=e/(2np+1)$.\cite{Wen95,Kane95,Ferraro08} This could explain the evolution of the effective charge with a  
crossover between the two more dominant 
excitations: the single-qp and the $|p|$-agglomerate. This mechanism, taking also into account of the non-universal value of the $\chi$LL 
exponents (environmental renormalizations)~\cite{Braggio12}, 
seems enough to explain the experiments both for co-propagating or counter-propagating edge states models.\cite{Ferraro10,Ferraro10b}\\
Lately a similar observation was also reported for $\nu=5/2$ where, further lowering the temperature, the effective carrier 
charge $e^*=e/4$ increases 
reaching the charge of $2$-agglomerates $e/2$.\cite{Dolev10} We have shown that, also for the case of a non-Abelian model, such 
as the anti-Pfaffian, in the 
presence of a renormalization mechanism for the bosonic modes, the agglomerates $e/2$
could dominate at low enough energies.\cite{DallaTorre10,Braggio12} We reported elsewhere a very good agreement with the observations using 
this approach.\cite{Carrega11}\\
The $\nu=5/2$ Pfaffian and anti-Pfaffian non-Abelian models correspond, respectively, to the $k$-level Read-Rezayi (RR) theory~\cite{Read99} and 
its particle-hole conjugate, the anti-Read Rezayi ($\overline{\textrm{RR}}$)~\cite{Levin07,Lee07}, 
with $k=2$. Consequently one is legitimate to ask if similar crossovers could be, in principle, also observed in the generic models 
based on $k$-level RR theories. Here, we will discuss this issue in detail finding that agglomerates may be dominant at low energies 
both for RR and $\overline{\textrm{RR}}$ models with some relevant exceptions.\cite{Read99,Bishara08} 

In particular we found that for $k>4$ the $2$-agglomerate dominates the transport in $k$-level 
RR model but, when the charged modes are renormalized by environmental effects, the single-qp could become again dominant. We also found, for RR, an unexpected result for $k=2,3$ where the single-qps are always the dominant even in the presence of renormalization effects.

For $k$-level $\overline{\textrm{RR}}$ models the situation is 
even more complex. Without any renormalization the single-qp and the $2$-agglomerate are equally dominant 
because they have exactly the same scaling dimensions for any $k$. 
We will see that charged mode renormalization, induced by the external environment, favors the dominance of the single-qp, while the neutral one 
helps the $2$-agglomerates. In the paper we precisely determine the conditions on the renormalization strengths where one excitation will dominate 
over the other. 

The rest of the paper is organized in three sections. In Sec.II we present the edge state models we will investigate: the RR and $\overline{\textrm{RR}}$. The peculiar algebraic properties 
of parafermions and the excitation structure of the RR states are described in Sec.~IIA. The peculiar forms of the excitations in terms of the 
composition of a parafermionic neutral sector and the charge bosonic sector are also discussed. In Sec.~IIB we investigate instead the excitation 
structure of the $\overline{\textrm{RR}}$ models in the disorder dominated limit\cite{Bishara08}. In Sec.~III we will discuss which are the 
dominant excitations of the two models by investigating the operator scaling dimension of the excitations. We will took also into account of the 
possible effects of renormalization induced by an the external environment. Finally in Sec.~IV we will conclude the discussion pointing out 
some consequences for the non-Abelian models of $\nu=5/2$.

\section{Edge states models}
For filling factor in the lowest Landau level (LL) the Laughlin and the hierarchical construction were quite successful. Unfortunately, for 
the fractional values in the second Landau level, such as 
$\nu=2+1/2,2+2/5,2+3/5,2+2/3,..$, many theories are not able to correctly describe the system just because 
the "vacuum" is now constituted by two filled LLs. Hereafter we consider 
two of the most successful proposals, discussed in the literature: the RR states and their particle-hole conjugate 
$\overline{\textrm{RR}}$, which are based on a non-Abelian extension of the fractional statistics.\cite{Stern09} 
One of the interest in these models is determined by their intrinsic  non-Abelian nature that is potentially relevant for topological
quantum computation.\cite{Nayak08}  

\subsection{Read-Rezayi models}
A serious step to go beyond the discussed pitfall was proposed by Read and 
Rezayi~\cite{Read99} that introduced 
a completely new class of wave-functions, a sort of generalization of the concept of Pfaffian state.\cite{Moore91} In particular it was
showed, using conformal field theory 
arguments for $k\geq2$,  that the  eigenstate of $k+1$-body $\delta$-function interaction can be written in terms of  a generalization of the 
Majorana fermions: the $\mathbb{Z}_k$ parafermions. These parafermions correspond to an 
$\mathrm{SU}(2)_k/\mathrm{U}(1)$  coset 
where the central 
charge is given by $c=(2k-2)/(k+2)$.\cite{Zamolodchikov85} The $k$-level RR state describe filling factor $\nu^*=k/(k+2)$ in terms of a charged and 
a neutral sector. 

The 
imaginary time action for the edge states in these models is~\cite{Read99,Bishara08} 
\begin{equation}
\label{eq:action}
\mathcal{S}=\int\!\! d\tau dx\  \left[\left(\frac{1}{4\pi\nu_\rho}\right)[\partial_x \varphi_\rho
(i\partial_\tau+v_\rho\partial_x)\varphi_\rho]+\mathcal{L}_k\right]\ , 
\end{equation}
  where $\varphi_\rho$ is the bosonic charged mode propagating at the velocity $v_\rho$ with $\nu_\rho=k/(k+2)$ and satisfying the commutation relation 
 \begin{equation}
[\varphi_\rho(x),\varphi_\rho(x')]=i\pi \nu_\rho \textrm{sgn}(x-x').
\end{equation}
The particle density operator on the edge can be written as 
 $\rho(x)=\partial_x \varphi_\rho/(2 \pi)$.
 The neutral sector $\mathcal{L}_k$ coincides with the SU(2) Weiss-Zumino-Witten model where the 
 $\mathrm{U}(1)$ has been gauged.\cite{Bishara08} For example in the case of $\nu=5/2=2+1/2$, when $k=2$, the neutral mode 
 is the Majorana fermion $\mbox{\boldmath $\varPsi$}$ of the Pfaffian state $\mathcal{L}_2=i\mbox{\boldmath $\varPsi$}(i\partial_\tau+v_\sigma\partial_x)
 \mbox{\boldmath $\varPsi$}$ propagating at velocity $v_\sigma$. 
 
 In general in the parafermionic neutral sector there are primary fields $\mathbf{\Phi}_{j,m}$ with the integer or half-integer number $j$ satisfying 
 $0\leq j\leq k/2$ and 
 $m\in(-j,-j+1,....,j)$. It is easy to see that $j$ and $m$ are both half-integer or integer (such as in the usual spin algebra where $j$ is the total 
 spin and $m$ is a projection along one axis). 
 The primary fields satisfy additional identities such as $\mathbf{\Phi}_{j,m}\equiv\mathbf{\Phi}_{j,m+k}$ and  
 $\mathbf{\Phi}_{j,m}\equiv\mathbf{\Phi}_{k/2-j,m+k/2}$ that 
 reduce the number of them to  $k(k+1)/2$.
 For example for $k=2$ we have three primary fields: the identity $\mathbf{\Phi}_{0,0}\equiv \mathbf{I}$, the twist 
 field  $\mathbf{\Phi}_{1/2,-1/2}\equiv \mbox{\boldmath $\sigma$}$ and the 
 Majorana fermion $\mathbf{\Phi}_{1,0}\equiv \mbox{\boldmath $\varPsi$}$.
 
 These operators $\mathbf{\Phi}_{j,m}$ have the $k$ dependent  conformal dimensions~\cite{Bishara08,Zamolodchikov85,Difrancesco97,Stern09,Cappelli10,Cappelli11} 
 \begin{equation}
 \label{delta}
\delta_{j,m,k}=\frac{j(j+1)}{k+2}-\frac{m^2}{k}
 \end{equation}
 and satisfy the fusion algebra
 \begin{equation}
 \mathbf{\Phi}_{j,m}\times\mathbf{\Phi}_{j',m'}\to\sum^{\min[j+j',k-j-j']}_{j''=|j-j'|} \mathbf{\Phi}_{j'',m+m'}
 \end{equation}
 derived from the operator product expansion.
 
 The fundamental charge for this class of states is $e^*=e/(k+2)$. Allowed qps  excitations can be 
 written in terms of the product of the neutral field operators $\mathbf{\Phi}_{j,m}$ and a standard vertex operator for the charge 
 sector $e^{i\alpha\varphi}$ with the prescription 
 \begin{equation}
 \label{PsiRR}
 \mathbf{\Psi}\propto\mathbf{\Phi}_{j,m}\ e^{i\alpha\varphi_\rho}
 \end{equation}
with the coefficient $\alpha$, that determines the charge of the excitation, given by
 \begin{equation}
[\rho(x),e^{i\alpha\varphi_\rho}]=-\alpha\nu_\rho\delta(x-x')\ e^{i\alpha\varphi_\rho}.
 \end{equation}
 The coefficients  $\alpha$, for all the possible excitations, can be determined by requiring the monodromy\footnote{Any excitation 
must acquire a trivial phase turning 
around an electron. The same electrons operators are identified on a the base of a 
self-consistent monodromy argument.} condition.\cite{Wen92,Read90,Froelich91,Ferraro10e} The 
unit charge operator, with fermionic statistics,  obtained with this approach is the electron operator
 \begin{equation}
\mathbf{\Psi}^{(e)}\propto\mathbf{\Phi}_{k/2,1-k/2} e^{i \varphi_\rho(k+2)/k}\ .
\end{equation}
The most general excitation is labelled by three numbers $(n,j,m)$ 
and is written as
 \begin{equation}
 \label{PsiRRgen}
\mathbf{\Psi}_{n,j,m}\propto \mathbf{\Phi}_{j,m}\  e^{i n\varphi_\rho/k} 
\end{equation}
where the integer $n=(u k+2m)$, with $u\in\mathbb{Z}$. 
The 
excitation charge $q_{n,j,m}=e\  n/(k+2)$  is an integer multiple of the minimal charge $e_k^*=e/(k+2)$. 
For $n=1$ we have the single-qp (minimal charge) and for $n\geq2$ the $n$-agglomerate of qps. For an even $k$ and 
half-integer (integer) $j$ the $n$-agglomerate charge must be an odd (even) multiple of the minimal charge $e_k^*$.

The single-qp with minimal charge $e_k^*$ is represented by the superpositions of operators according to  
 \begin{equation}
\mathbf{\Psi}^{(qp)}\propto\sum_{m=\pm1/2}\gamma_m\ \mathbf{\Phi}_{1/2,m}\  e^{i \varphi_\rho/k} 
\end{equation}
where $\gamma_m$ are arbitrary coefficients.\cite{Difrancesco97,Stern09} 

For example, 
for $\nu=5/2$,  the qp operator is $\mathbf{\Psi}^{(qp)}\propto\mathbf{\Phi}_{1/2,\pm1/2}\  e^{i \varphi_\rho/2}\equiv\mbox{\boldmath $\sigma$}e^{i\varphi_\rho/2}$ with 
charge $e^*=e/4$. 
Recently experimental observations confirmed the presence of this quarter of an electron charge giving a 
strong support to models based 
on Read-Rezayi states for this fraction.\cite{Radu08,Bid10,Dolev08,Venkatachalam11}

\subsection{The particle-hole conjugate Read-Rezayi models}
In full analogy with the anti-Pfaffian state, for $\nu=5/2$, Bishara et al.~\cite{Bishara08} introduced the $k$-level particle-hole conjugate 
Read-Rezayi ($\overline{\textrm{RR}}$) model. The edges, in these models, are composed by one filled LL~\footnote{The two lowest LLs are 
again the "vacuum" of 
the theory and will be neglected.} and a $k$-level RR state of holes superimposed on it. Then the bulk filling 
factor is $\nu=2+\nu^*$ with  $\nu^*=1-k/(k+2)=2/(k+2)$.

Hereafter we present only the theory for $\overline{\textrm{RR}}$ at the fixed point of the disorder dominated 
phase because, only for such limit, the appropriate value of quantized conductance is properly obtained.\cite{Kane95,Levin07,Lee07,Bishara08}
 At this fixed point one can write the Lagrangian in terms of a bosonic charged mode 
$\varphi_\rho$ with interaction parameter $\nu_\rho=2/(k+2)$ and propagating velocity $v_\rho$ and a counter-propagating 
neutral sector. This sector is composed by two modes with the 
same velocity $v_\sigma$: one bosonic mode $\varphi_\sigma$, with interaction parameter $\nu_\sigma=2/k$, 
and a $k$-level parafermion with Lagrangian $\mathcal{L}_k$. The low energy effective Lagrangian for the edge is~\cite{Bishara08} 
\begin{eqnarray} 
\mathcal{L}= &&\frac{1}{4\pi\nu_\rho} \partial_{x}\varphi_{\rho} \left(i\partial_{\tau} +v_{\rho} \partial_{x} \right)\varphi_{\rho}\nonumber \\
&&+\frac{1}{4\pi\nu_\sigma} \partial_{x}\varphi_{\sigma} \left(-i\partial_{\tau} +v_{\sigma} \partial_{x} \right)\varphi_{\sigma} + \mathcal{L}_k 
\end{eqnarray} 
where the commutation relations of the fields are 
 \begin{equation}
[\varphi_j(x),\varphi_{j'}(x')]=i \pi\delta_{j,j'} \xi_j \nu_j \textrm{sgn}(x-x') 
\end{equation}
with $j=\rho,\sigma$ and 
where $\xi_\rho= +1$ ($\xi_\sigma= -1$) indicates the downstream (upstream) mode propagation. The electron operator is given by the $m$-multiplet
superposition of the operators $\mathbf{\Phi}_{k/2,m}$ according to~\footnote{The bar notation over the operators, such as $\overline{\mathbf{\Psi}}$, it is used to distinguish  $\overline{\textrm{RR}}$ operators from the RR case discussed before.}
\begin{equation}
\label{eRRb}
\overline{\mathbf{\Psi}}^{(e)}\propto\sum_{m=-k/2}^{k/2}\gamma'_m\  \mathbf{\Phi}_{k/2,m}\ e^{i m \varphi_\sigma} e^{i \varphi_\rho(k+2)/2}, 
\end{equation}
with $\gamma'_m$ arbitrary coefficients. All the admissible excitations of the theory are calculated by applying the 
monodromy condition over these electron operators. Finally the generic allowed qp excitation is labeled by three numbers $(n,j,m)$
 \begin{equation}
 \label{PsiRRb}
 \mathbf{\overline{\Psi}}_{n,j,m}\propto  \mathbf{\Phi}_{j,m}\ e^{i m \varphi_\sigma}\ 
e^{i n \varphi_\rho/2} \end{equation}
 where  $n$ assumes even (odd) values  when $j$ is integer (half-integer). 
 The charge of the generic excitations of equation (\ref{PsiRRb}) is 
 $q_{n,j,m}=n e/(k+2)=ne_k^*$, an integer multiple of the minimal charge $e_k^*=e/(k+2)$. The independency of the charge from $m$ 
 assumes that the operator of an excitation with fixed charge $ne_k^*$, and angular momentum $j$, is given by a $m$-multiplet superposition
\begin{equation}
\overline{\mathbf{\Psi}}_{n,j}\propto\sum_{m=-j,-j+1,...}^{j}\gamma''_m\  \mathbf{\Phi}_{j,m}\ e^{i m \varphi_\sigma} e^{i n \varphi_\rho/2}, 
\end{equation}
 with $\gamma''_m$ arbitrary coefficients.

\section{Single-qp. vs agglomerate dominance}
In the following we will discuss which are, at low energies, the dominant excitations in the RR and $\overline{\textrm{RR}}$ models. This can be done by looking
at the long-time behaviour at $T=0$ of the imaginary time two-point Green's function $\langle T_{\tau}\Psi(\tau) \Psi^\dagger(0)\rangle
\propto |\tau|^{-2\Delta}$~\cite{Kane92} for the general qp operators. Let's start to consider the RR states.

\subsection{Read-Rezayi states}
The total scaling dimension for the RR states of a generic $n$-agglomerate operator $\mathbf{\Psi}_{n,j,m}$ of equation (\ref{PsiRRgen}) is 
given by 
\begin{equation}
\label{DeltaRR}
\Delta^{(RR)}_{n,j,m}=\frac{g_\rho}{2}\left(\frac{n^2}{k (k+2)}\right)+\frac{j(j+1)}{k+2}-\frac{m^2}{k}
\end{equation}
where the first term is the charge contribution and the other terms come from the neutral parafermionic sector (cf. conformal dimension of equation (\ref{delta})).

 In the previous formula we assumed 
that, in general, the charge sector could also be "renormalized" by the presence of interactions with the external environments with a factor 
$g_\rho\geq1$. Many mechanisms can cause such renormalization effects such as the coupling with phonons~\cite{Rosenow02}, dissipation induced 
by electromagnetic environment~\cite{CastroNeto97,Safi04} or the combined effect of out of equilibrium $1/f$ noise and 
dissipation~\cite{DallaTorre10,Braggio12}. We do not know any mechanisms acting on the parafermions  
and consequently we do not assume any renormalization for the scaling dimension in this sector.

In general, from the previous formula, and from the structure of operators in the RR theory given in equation (\ref{PsiRRgen}), 
one can see which is the most dominant excitation of the theory (i.e. the excitation operator with minimal scaling dimension). 

We found that the $n=1$ single-qp $\mathbf{\Psi}_{1,1/2,\pm1/2}$ (with charge $e^*_k$) is the dominant excitation for odd $n$, while the $2$-agglomerate $\mathbf{\Psi}_{2,0,0}$ (with charge $2e_k^*$) dominates for even $n$.\footnote{Note 
that for $k=3$ one could also imagine the presence of the single-qp $\mathbf{\Psi}_{1,1,\pm1}$ 
but due to the parafermionic identification property $\mathbf{\Phi}_{j,m}\equiv\mathbf{\Phi}_{k/2-j,m+k/2}$ one
demonstrates it \emph{coincides} with the single-qp $\mathbf{\Psi}_{1,1,\pm1}\equiv\mathbf{\Psi}_{1,1/2,\mp1/2}$.} 
The single-qp, based on the primary fields $ \mathbf{\Phi}_{1/2,\pm1/2}$, has not trivial fusion rules in the parafermionic sector and presents  
a non-Abelian statistics. The $2$-agglomerate, based on the identity operator  $\mathbf{\Phi}_{0,0}\equiv\mathbf{I}$, are instead an 
Abelian excitation. Between these two excitations one has to find which is the most relevant by comparing directly their scaling dimensions
 $\Delta^{(RR)}_{1,1/2,\pm1/2}$ and $\Delta^{(RR)}_{2,0,0}$.

In particular the single-qp  $\mathbf{\Psi}_{1,1/2,\pm1/2}$ always dominates for $k=2,3$. 
For $k=4$ the $2$-agglomerate has the same scaling of the single-qp and only for $k>4$ the $2$-agglomerate becomes more relevant.

If we take into account an environmental renormalization $g_\rho\geq1$, that acts only on the charged modes,
the single-qp is typically favored and, in general, always dominates 
when the renormalization is strong enough $g_\rho>(k-1)/3$.\footnote{This inequalities, with the 
additional condition for renormalization $g_\rho\geq 1$, shows that for $k=2,3$ the sigle-qp is indeed always dominating.}

\subsection{Anti Read-Rezayi states}
Let's consider now the $\overline{\textrm{RR}}$  states. The  $n$-agglomerate operator, given in equation (\ref{PsiRRb}), has the scaling 
dimension similar to the previous case but now we have an additional 
contribution from a bosonic neutral mode $\varphi_\sigma$. The scaling dimension becomes
\begin{equation}
\label{DeltaRRb}
\Delta^{(\overline{RR})}_{n,j,m}=\frac{g_\rho}{4} \frac{n^2}{k+2}+\frac{j(j+1)}{k+2}+\frac{m^2}{k}\left(g_\sigma-1\right)
\end{equation}
where we took into account the renormalization factors $g_\rho$ and $g_\sigma$ of the charged and neutral bosonic modes 
respectively. 
We discussed in Ref.\cite{Braggio12} the mechanisms that could determine the renormalization of the neutral modes with 
the restriction $g_\sigma\geq 1$, similar to the charge modes.\footnote{In Ref.\cite{Braggio12} was also demonstrated 
that the renormalization mechanism, based 
on the interplay between noise and dissipation, is also valid in the disorder dominated phase as required for $\overline{\textrm{RR}}$ models}  
It is important 
to note that neutral mode renormalizations can also be stronger than the charged one when $g_\sigma>g_\rho$.\cite{Braggio12} In conclusion
hereafter we treat $g_\rho,g_\sigma\geq 1$ as completely independent parameters. 

The 
contribution to the scaling dimension of the bosonic neutral mode component $e^{im\varphi_\sigma}$ alone is $g_\sigma m^2/k$, and it is 
added to the parafermionic sector contribution of equation (\ref{DeltaRR}) giving the result reported in equation (\ref{DeltaRRb}). Obviously 
one has also to take into account that the charge sector has a different $\nu_\rho$ with respect to the RR model. 
All these differences contribute to create the peculiar behaviour of the $\overline{\textrm{RR}}$ model described hereafter. 

Indeed we firstly observe that, in absence of 
renormalizations ($g_\rho=g_\sigma=1$), the term proportional to $m^2$ vanishes. The most dominant 
excitations are of two classes: the non-Abelian single-qp 
$\overline{\mathbf{\Psi}}_{1,1/2,\pm1/2}$, with minimal charge $e_k^*=e/(k+2)$, and the Abelian $2$-agglomerate  
$\overline{\mathbf{\Psi}}_{2,0,0}$, with 
charge $2e^*_k$. These excitations have equal scaling $\Delta^{(\overline{RR})}_{1,1/2,\pm1/2}=\Delta^{(\overline{RR})}_{2,0,0}=1/(k+2)$. 

If, instead, we assume a renormalization of 
the bosonic modes ($g_\rho,g_\sigma>1$) depending on the precise values of the parameters one type of excitation will dominate over the other. 

Here we simply calculate the condition to have the dominance of the $2$-agglomerate over the single-qp 
$\Delta^{(\overline{RR})}_{2,0,0}<\Delta^{(\overline{RR})}_{1,1/2,\pm1/2}$. This leads to the general relation between the renormalization parameters
\begin{equation}
\label{ineq}
g_\rho<\left[ 1+\frac{(k+2)(g_\sigma-1)}{3k}\right],
\end{equation}
for which the $2$-agglomerate is dominant over the single-qp.

For example, if charged modes are not renormalized ($g_\rho=1$) and instead the neutral modes are ($g_\sigma>1$) the 
dominance of the $2$-agglomerates is guaranteed. The opposite happens when $g_\sigma=1$ and $g_\rho\geq1$ where indeed 
the single-qps dominate. 

In general charged renormalization favors the 
dominance of the single-qp (in force of their smaller charge) while the neutral mode renormalization leads to the dominance of 
the $2$-agglomerate ( because it has not neutral contribution $j=m=0$). 

In conclusion to determine which excitation will be dominant we need a precise knowledge of non-universal renormalization parameters 
$g_\rho$ and $g_\sigma$. These parameter can be, in principle, deduced by the fitting of the transport properties and crosschecking,  a posteriori,
if the excitation observed to be dominant coincide with these theoretical predictions. 

For example, we recently considered the anti-Pfaffian case of $5/2$ where 
a comparison with experimental observation~\cite{Dolev10, Carrega11} is possible. In such a case, the renormalization parameters 
can be extracted by looking at the scaling of the transport properties because the power laws are also directly affected 
by the renormalization parameters. 
We found that the dominance of the $2$-agglomerate at low temperatures is predicted in fully agreement with the back-scattering 
conductance and QPC noise properties.
The same experiment consistently indicates that at higher temperatures (higher energies) the dominant charge evolves 
from the $2$-agglomerate to the single-qp.
This behavior is observed because the single-qp has higher scaling dimension with respect to the $2$-agglomerate. So, by increasing temperatures,
single-qp could naturally overcome the agglomerate contribution. This example shows how the comparison of the transport and noise properties 
in QPC setup in the weak-backscattering could validate some of the results described here. 

\section{Conclusions}
\label{sec:Conclusions}
In conclusions we demonstrated that, in the $k$-level RR states, the single-qp is \emph{always} the dominant excitation for $k=2,3$. In 
the presence of sufficient strong renormalization of the charge sector, $g_\rho>(k-1)/3$, the single-qp dominance is guaranteed for all $k$-level RR models.

For the $\overline{\textrm{RR}}$ states, instead, the single-qp and the $2$-agglomerate are equally relevant. In the presence of 
renormalization effects, the single-qp dominance is favored by renormalizations of the charge modes while
the $2$-agglomerate dominance is favored by neutral renormalizations.

Finally we note that the observation of the dominance of a $2$-agglomerate or, even, the crossover between this excitation 
and the single-qp,  in the backscattering conductance and noise transport in QPCs, could provide an indication toward the applicability 
of the $k$-level RR or $\overline{\textrm{RR}}$ models 
to some specific Hall states in the second Landau level. 

In particular our analysis may be relevant for $\nu=5/2$ where the agglomerates was recently seen.\cite{Dolev10} According to our results this would  
indicate that Pfaffian ($2$-level RR state) mode is probably excluded because predict the dominance of the single-qp independently 
by any renormalization effect).\footnote{This conclusion is deeply associated with the reasonable assumption that the conformal dimension of the  
parafermionic sector cannot be modified by any external environment. If a similar mechanism is found this conclusion has to be necessary modified.}    
Furthermore the good agreement of the anti-Pfaffian with the intriguing observation of a neutral counter-propagating 
mode~\cite{Bid09} and various transport properties~\cite{Radu08,Dolev10,Carrega11} support the appropriateness of the second model. Anyway 
the discussion is still very debated.\cite{Lin12} 

Certainly the observation of the dominance of agglomerates at low temperatures 
would be probably possible for many models including the two most important non-Abelian sequence such as RR and $\overline{\textrm{RR}}$. 
Conversely the same observation of the dominance of an agglomerate could returns interesting information on the model nature.

\begin{acknowledgments}
We thank M. Sassetti, G. Viola and M. Carrega for valuable discussions and acknowledge the support of the CNR STM 2010 program and the 
EU-FP7 via ITN-2008-234970 NANOCTM.
\end{acknowledgments}


\begin{thebibliography}{10}
\bibitem{Laughlin83}{Laughlin R B 1983 \PRL~\textbf{50} 1395}
\bibitem{dePicciotto97}{De Picciotto, Reznikov M, Heiblium M, Umansky V, Bunin G and Mahalu D 1997 \textit{Nature}  \textbf{389} 162}
\bibitem{Saminadayar97}{Saminadayar L, Galttli D C, Jin Y and Etienne B 1997 \PRL~\textbf{79} 2526}
\bibitem{Wen91}{Wen X G 1991 \PRL~\textbf{66} 802}
\bibitem{Read99}{Read N and Rezayi E 1999 \textit{Phys. Rev. B} \textbf{59}  8084}
\bibitem{Fendley07}{Fendley P, Fisher M P and Nayak C 2007 \textit{Phys. Rev. B}  \textbf{75} 045317}
\bibitem{Levin07}{Levin M, Halperin B I and Rosenow B 2007 \PRL~\textbf{99} 236806} 
\bibitem{Lee07}{Lee S S, Ryu S, Nayak C and Fisher M P A 2007 \PRL~ \textbf{99} 236807}
\bibitem{Chang03}{Chang A M 2003 \textit{Rev. Mod. Phys.} \textbf{75} 1449}
\bibitem{Wen90}{Wen X G 1990 \PRL~ \textbf{64} 2206}
\bibitem{Jain89}{Jain J K 1989 \PRL~ \textbf{63} 199}
\bibitem{Wen92}{Wen X G and Zee A 1992 \textit{Phys. Rev. B} \textbf{46} 2290}
\bibitem{Kane94}{Kane C L and Fisher M P A 1994 \PRL~ \textbf{72} 724}
\bibitem{Reznikov99}{Reznikov M, De Picciotto R, Griffiths T G, Heiblum M and Umansky V 1997 \textit{Nature} \textbf{399} 238}
\bibitem{Chung03}{Chung Y C, Heiblum M and Umansky V 2003 \PRL~ \textbf{91} 216804}
\bibitem{Bid09}{Bid A, Ofek N, Umansky V and Heiblum M  2009 \PRL~ \textbf{103}  236802}
\bibitem{Wen95}{Wen X G 1995 \textit{Adv. Phys.} \textbf{44} 405} 
\bibitem{Kane95}{Kane C L and Fisher M P A 1995 \textit{Phys. Rev B} \textbf{51} 13449}
\bibitem{Ferraro08}{Ferraro D, Braggio A, Merlo M, Magnoli N and Sassetti M 2008 \PRL~ \textbf{101} 166805}
\bibitem{Braggio12} Braggio A, Ferraro D, Carrega M, Magnoli N and Sassetti M \textit{(Preprint arXiv:1203.1906)}
\bibitem{Ferraro10}{Ferraro D, Braggio A, Magnoli N and Sassetti M 2010 \textit{New J. Phys} \textbf{12} 013012}
\bibitem{Ferraro10b}{Ferraro D, Braggio A, Magnoli N and Sassetti M 2010 \textit{Phys. Rev. B} \textbf{82} 085323}
\bibitem{Dolev10}{Dolev M, Gross Y, Chung Y C, Heiblum M, Umansky V and Mahalu D 2010 \textit{Phys. Rev. B} \textbf{81} 161303R}
\bibitem{DallaTorre10}{ Dalla Torre E G, Demler E, Giamarchi T and Altman E  2010 \textit{Nature Phys.} \textbf{6} 806}
\bibitem{Carrega11}{Carrega M, Ferraro D, Braggio A, Magnoli N and Sassetti M 2011 \PRL~ \textbf{107} 146404}
\bibitem{Bishara08}{Bishara W, Fiete G A and Nayak C 2008 \textit{Phys. Rev. B} \textbf{77} 241306}
\bibitem{Stern09} {Stern A 2008 \textit{Ann. Phys.} \textbf{323} 204}
\bibitem{Nayak08}{Nayak C, Simon S H, Stern A, Freedman M, and Das Sarma S 2008 \textit{Rev. Mod. Phys.} \textbf{80} 1083}
\bibitem{Moore91}{Moore G and Read N 1991 \textit{Nucl. Phys. B} \textbf{360} 362}
\bibitem{Zamolodchikov85}{Zamolodchikov A B and Fateev V A 1985 \textit{Sov. Phys. JETP} \textbf{62} 215}
\bibitem{Difrancesco97}{Di Francesco P, Mathieu P and S\'en\'echal D \textit{Conformal field theory}, Springer-Verlag, New York 1997}
\bibitem{Cappelli10} {Cappelli A, Viola G, Zemba G R 2010 \textit{Ann. Phys.} \textbf{325} 465}
\bibitem{Cappelli11} {Cappelli A, Viola G 2011 \textit{J. Phys. A} \textbf{44} 075401}
\bibitem{Read90}{Read N 1990 \PRL~ \textbf{65} 1502}
\bibitem{Froelich91}{Fr\"oelich J and Zee A 1991 \textit{Nucl. Phys. B} \textbf{364} 517}
\bibitem{Ferraro10e}{Ferraro D, Braggio A, Magnoli N and Sassetti M 2010 \textit{Physica E} \textbf{42} 580}
\bibitem{Radu08}{Radu I P, Miller J B, Marcus C M, Kastner M A, Pfeiffer L N and West K W 2008 \textit{Science} \textbf{16} 899}
\bibitem{Bid10}Bid A, Ofek N, Inoue H, Heiblum M, Kane C L, Umansky V and Mahalu D 2010 \textit{Nature} \textbf{466} 585
\bibitem{Dolev08} Dolev M, Heiblum M, Umansky V, Stern A and Mahalu D 2008 \textit{Nature} \textbf{452} 829
\bibitem{Venkatachalam11} Venkatachalam V, Yacoby A, Pfeiffer L and West K  2011 \textit{Nature} \textbf{469} 185
\bibitem{Kane92}{Kane C L 1992 \PRL~ \textbf{68} 1220}
\bibitem{Rosenow02}{Rosenow B and Halperin B I 2002 \PRL~ \textbf{88}  096404}
\bibitem{CastroNeto97}{CastroNeto H, Chamon C de C and Nayak C 1997 \PRL~ \textbf{79} 4629}
\bibitem{Safi04}{Safi I and Saleur H 2004 \PRL~ \textbf{93} 126602}
\bibitem{Lin12}{Lin X, Dillard C, Kastner M A , Pfeiffer L N, 
West K W, \textit{(Preprint arXiv:1201.3648)}}
\end{thebibliography}
\end{document}